# Open Government Data in Russian Federation

Dmitrij Koznov[1], Olga Andreeva[1], Uolevi Nikula[2], Andrey Maglyas[2], Dmitry Muromtsev[3], Irina Radchenko[3]

[1] *Software Engineering Department, Saint Petersburg State University, Bibliotechnaya sq., 2, Saint Petersburg, Russia*
d.koznov@spbu.ru, aovladi@gmail.com

[2] *Department of Innovation and Software, Lappeenranta University of Technology, Skinnarilankatu, 34, Lappeenranta, Finland* { uolevi.nikula, Andrey.Maglyas}@lut.fi

[3] *Department of Information Systems, ITMO University, Kronverkskiy pr., 49, Saint Petersburg, Russia*
d.muromtsev@gmail.com, iradche@gmail.com

Abstract: Open data can increase transparency and accountability of a country government, leading to free information sharing and stimulation of new innovations. This paper analyses government open data policy as well as open data initiatives and trends in Russian Federation. The OCDE analytical framework for national open government data portals and supporting initiatives is used as the bases for the study. The key issues of Russian open government data movement are summarized and aggregated. The paper argues the necessity of systematic development of the open data ecosystem, the leading role of the government in data release, a deeper business involvement, and a reduction of bureaucratic barriers.

## 1 INTRODUCTION

Open Government (OG) is a movement to make government activities more open and transparent, with open government data (OGD) as its essential part (Gigler, Custer, Rahmetulla, 2011). Today governments produce huge amounts of information through their various activities, from statistics used in policy-making to budget and financial auditing data. Taxpayer-funded production and opening of this information may bring into being public-private communities to develop innovations outside the government.

There are many international initiatives around OGD at the moment, such as Open Government Partnership, Open Knowledge Foundation, and Open Data Institute. Many countries have introduced open data portals, and according to (Hendler, Holm, Musialek, Thomas, 2012), the most advanced national portals are those in the U.S., the U.K., France and Singapore. A number of new information technologies have appeared to support the development of open data, portals, and applications, e.g. CKAN (http://ckan.org/) and Socrata (http://www.socrata.com/). All this has brought up new research topics and questions like access to data, accountability, coordination mechanisms for open data activities, data sharing, information and knowledge sharing (Zuiderwijk, Helbig, Gil-García, Janssen, 2014).

Open data ratings like the Open Global Data Index (OGDI) (OGDI, 2015) and the Open Data Barometer (ODB, 2015) have been developed to provide a quick overview on open data across the globe. However, these ratings only offer quantitative comparative surveys to give an overview of OGD in a large number of countries, but not to provide detailed information about each country. Consequently, there are many research papers about national open data: the U.S. (Hendler, Holm, Musialek, Thomas, 2012), the U.K. (Shadbolt, et al., 2012), Canada (Roy, 2014), Brazil (Albano, Reinhard, 2014), Mexico (González, Garcia, Cortés, Carpy, 2014), India (Agrawal, Deshmukh, Srinivasa, etc. 2013), Greece (Alexopoulos, Spiliotopoulou, Charalabidis, 2013), Latvia (Bojars, Liepins, 2014), Kenya (Mutuku, Colaco, 2012), etc.

Such studies analyze different OGD trends, present innovations and successful experiences, and thereby provide a basis for knowledge and information sharing in the OGD community around the world.

OGD movement in Russian Federation (R.F.) started in 2012, when the first national OGD concept was developed. In 2013, the R.F. with other members of the G8 group approved Open Data Charter (G8, 2013) to facilitate progress in OGD and international collaboration. In 2006-2015 a number of laws and regulations were issued in the R.F. to support government information sharing. In 2014, the R.F. OGD Portal (http://data.gov.ru/) was launched, and at the moment more than 7500 dataset have been published there.

There are several documents and reports about open data in the R.F. (HSE, 2012), (Zhulin, 2013), (Castro, Korte, 2015), (OGD Recommendations, 2014), (Russian OGD Plan, 2014), (Begtin, Vovk, Sakoyan, 2015). Even though some of them have been translated into English, most of them are in the R.F. Other limitations with the available information is that they focus on particular aspects of OGD in the R.F., often lack analytical basis, are often incomplete, and more generally we failed to find research publications about OGD in Russia. To address these shortcomings, the current paper studies OGD in the R.F. using an analytical framework for national OGD portals and supporting initiatives from (Ubaldi, 2013).

This paper is structured as follows. Section 2 offers a brief review of the OGD movement, and Section 3 reviews literature on OGD in the R.F. Section 4 discusses the research methodology. Section 5 contains the results, including the challenges of OGD in the R.F., while Sections 6 deals with the conclusions.

## 2 BACKGROUND

### 2.1 Open data movements in different countries

The issue of free access to government information is an important element of democracy, for example, it was mentioned in «The Freedom of Information Act» passed in 1967 in the U.S. The Act allows for full or partial disclosure of previously unreleased information and documents controlled by the United States government. In 1978, France enacted its Law on Free Access to Administrative Documents (Law No. 78-753) and in 2003, the European Union implemented Directive 2003/98/EC on the reuse of public sector information.

The number of national OGD initiatives has increased steadily since 2009. Along with more economically developed countries such as the U.S., the U.K., and France, OGD is rapidly evolving in developing countries like Kenya and Ghana. One report groups countries into three categories with regard to open data development level (Capgemini, 2013): trendsetters (the U.S., the U.K., France, Canada, Australia), followers (Denmark, New Zealand, Singapore, Belgium, Italy, Spain, Moldova, Ghana, Kenya, Chile, Norway, Hong Kong), and beginners (Austria, Estonia, Saudi Arabia, the UAE, Morocco). As of 2015, according to the OGDI (OGDI, 2015), the top ten OGD countries are Taiwan, the U.K., Australia, Denmark, Colombia, Finland, Uruguay, the U.S., the Netherlands, and France. The Open Data Barometer (ODB, 2015) lists the following countries as leaders in 2015: the U.K., the U.S., Sweden, France, New Zealand, the Netherlands, Norway, Canada, Denmark, and Australia. The two sources thus agree that the U.S. and the U.K. are leading OGD countries. In this section, we will consider briefly their open data movement.

### 2.1.1 United State

In 2009, the U.S. government launched the national open data portal to serve as a platform for U.S. federal agencies to publish data for public access. In May 2012, a digital government strategy was released, with OGD as an important facilitator of e-Government development. In 2013, an executive order made open and machine-readable format the new default for government information. As of today, the U.S. portal of open data contains approximately 159 000 datasets published by 77 organizations. The most considerable contributors are the National Oceanic and Atmospheric Administration of the Department of Commerce (71 945 datasets), the National Aeronautics and Space Administration (32071 datasets), the U.S. Fish and Wildlife Service, the Department of the Interior (26 401 datasets).

### 2.1.2 United Kingdom

The U.K. government launched its open data portal in 2010 and published the Open Government License for public sector information the same year. In 2012, the Open Data Institute was established to promote the use of government open data in the U.K. The U.K. government open data portal currently contains 31 150 datasets published by ten organizations. The most important contributors are the U.K. Hydrographic Office (4 039 datasets), the Environment Agency (2 154 datasets), Natural England (1 622 datasets), and the Office for National Statistics (1 435 datasets).

### 2.1.3 Summary

It should be noted that it is the countries with developed economies and ICT infrastructure that have the most advanced OGD. Another crucial issue for success of OGD movement is the mature civil society. At the same time, the situation with OGD varies significantly from country to country, including the leading ones: there are considerable differences even in quantitative rates, such as the numbers of datasets or contributors. Besides, there is an ambiguity in assessment that is becoming increasingly obvious: numerous ratings and studies significantly differ in their positioning of various countries, which makes the research of national OGDs essential and timely.

## 2.2 Open Data Charter

In 2013, all G8 countries signed an Open Data Charter which was an agreement designed to advance open data within these countries (G8, 2013). This document is a set of principles, which create a foundation for access to data as well as for release and reuse of data. The principles include the following: Open by Default, Timely and Comprehensive, Accessible and Usable, Comparable and Interoperable, For Improved Governance and Citizen Engagement, For Inclusive Development and Innovation. The Charter is one of the main starting points of OGD development in the R.F.

## 2.3 Value Creation

It is necessary to clear identified *value* to be created through OGD initiatives. Otherwise it is not possible to provide effectiveness of the OGD movement. Based on (Ubaldi, 2013) there are three value types which are economic, political, and social value. Economic value can be created by encouraging new firm creation, developing new services and products improving existing ones, generating increased tax incomes (Capgemini, 2013). Political value can be created by increasing the transparency of government processes and providing democratic control. Social value can be created by improving social services, building the next generation of empowered civil servants (Millard, 2013).

## 2.4 Public Sector and OGD Ecosystem

Public sector is the largest source of open data. According to (Investorwords, 2016), *public sector* is a part of the economy concerned with providing basic government services. The composition of the public sector varies by country, but in most countries the public sector includes such services as the police, military, public roads, public transit, primary education and healthcare..

*OGD ecosystem* (further – ecosystem) is a community of key actors of OGD initiatives on national/subnational levels (Ubaldi, 2013). Establishment of the right ecosystem means the involvement of various categories of actors and the provision of business support and stimulation of OGD usage. Constructing an ecosystem is necessary since otherwise OGD movement cannot be sustainable and socially beneficial (Janssen, Charalabidis, Zuiderwijk, 2012).

## 3. RELATED WORKS

The Higher School of Economics (one of the leading Russian universities) surveyed in 2012 the largest Russian open data beneficiaries (business companies, mass media, non-government organizations (NGOs), experts and bloggers), and tried to determine which government data should be opened first, and how it can be used by businesses (HSE, 2012). The survey revealed numerous problems such as lack of a uniform policy and standards in open data movement in the R.F.; no courses to train government staff to work with open data; an absence of necessary attributes in many datasets; and a lack of convenient services for information search. The survey also identified which data the business community is most interested in: various government registries, transport data, geo & map data, construction data, etc. Another result of the survey was a list of ideas for new services based on open data.

Another report describes the OGD situation in the R.F. in 2013 (Zhulin, 2013). This report reviews the primary laws and regulations that apply to open data, looks upon government bodies' open data portals, and outlines public initiatives.

A third report focuses on the progress of the G8 countries towards the principles of the Open Data Charter (Castro, Korte, 2015). The progress was scored based on how well each principle of the Charter was met, the total maximum being 100 points. The countries received the following ranks: the United Kingdom (90 points), Canada (80 points), the United States (80 points), France (65 points), Italy (35 points), Japan (30 points), Germany (25 points), and the R.F. (5 points). The 5 points result for the R.F. seems debatable, however. Namely, the five points were granted for licensing on the Data Portal of the R.F., which in fact is one of the weakest aspects of the Russian open data. On the other hand, the report points out that public access to government information is not backed by appropriate and sufficient legislation. Still the Federal Law No 112-FZ (adopted in June 2013) addresses this issue even though the terms of the law might be more consistent with the open data definition given in (Open Definition, 2016). Yet, the conclusions of the report are of great importance for the future progress of the open data movement in the R.F., i.e. for facilitating the partnership between the government, the civil society, and the private sector to prioritize data releases, for providing more support of innovation in open data, for raising the quality of open data, and for making data open by default both legally and in practice.

Finally, the Russian NGO Information Culture published a report in 2015 on the results of the government work towards opening key datasets in the R.F. as well as discusses the major projects in OGD (Begtin, Vovk, Sakoyan, 2015). Also, it analyses the legal background of OGD regulation in the R.F.

To summarize, the reports considered above provide typically collections of facts than studies on the given topics. The lack of methodological support results in incomplete information and difficulties with conclusions and recommendations. We failed to find a paper that would treat the situation in the R.F. systematically. Besides, most of the reports discussed above are written in the Russian language.

## 4. METHODOLOGY

The study of OGD initiatives in different countries runs into a number of difficulties because of differences in government organization, legislation, information culture, business involvement in OGD consumption (Erickson, Viswanathan, Shinavier, 2013). One can see that papers on various national OGD use informal and narrative approaches (Hendler, Holm, Musialek, Thomas, 2012), (Shadbolt, et al., 2012), (Roy, 2014), or some particular criteria: the country's geographic coverage by its open data (Agrawal, Deshmukh, Srinivasa, etc. 2013), the technologies used for portal development (Alexopoulos, Spiliotopoulou, Charalabidis, 2013), (Mutuku, Colaco, 2012), the assessment of metadata (Alexopoulos, Spiliotopoulou, Charalabidis, 2013), a number of datasets in different data categories (Agrawal, Deshmukh, Srinivasa, etc. 2013), (Bojars, Liepins, 2014), OGD formats analysis (Agrawal, Deshmukh, Srinivasa, etc. 2013), (Alexopoulos, Spiliotopoulou, Charalabidis, 2013). Actually, in every survey a special methodology is created. But due to a wide range of conditions and priorities in different countries these methodologies are hard to reuse. The same problem is for methodologies, created in the context of the special research task, e.g. see (Castro, Korte, 2015), (Capgemini, 2013).

To deal with the problem of unification OGD initiatives in various countries Open Global Data Index (OGDI, 2015), and the Open Data Barometer (ODB, 2015) have been developed. They are based on relatively simple quantitative metrics that make it possible to calculate the numeral index for each country and to create a global census. With the advantages of this approach, it fails at assessing OGD in terms of society and country information landscape, considering only OGD themselves.

For our research we have chosen an analytical framework for national OGD portals and supporting initiatives suggested in (Ubaldi, 2013). In contrast methods mentioned above, it focuses on OGD in connection with different aspects of the government/society issues and creation value. The framework allows to assess national OGD strategy, legal, technical, organization, communication and interaction issues, impact OGD on economy, political and social value creation. Below each component of the framework is briefly described below.

Overarching issues:
- Overall vision: overall strategy and the priorities of national OGD initiatives, connection between OGD and OG, coordination of OGD initiatives between central and local level.
- Governance/institutional framework: institutional supporting data development; accountability and responsibility frameworks.

Implementation:
- Legal framework and policy environment.
- Technical issues focus on technical matters that sustain or limit real data openness (data quality, interoperability, workflow for data

- release and approval, dataset storage, data cataloguing and metadata).
- Economic and financial: business case model, financing mechanisms to sustain the OGD portal, ensuring value creation for the whole economy and society.
- Organizational issues focus on the measure taken to enable and foster the changes required in the public sector: measures in place to ensure accountability, quality of data, etc.; measures to shift the culture of the public sector towards OGD; initiatives to ensure "buying in" of all stakeholders within the public sector.
- Communication and interaction: establishment of an OGD ecosystem, including measures to increase public interest in OGD, to provide feedback, etc.

Impact:
- Measures and mechanisms to appraise the impact of OGD initiatives on economic, political and social value creation, monitoring user satisfaction, percentage of datasets released for a specific purpose, etc.

Our study uses the following reports and documents as information sources: the Bulletin of the Information and Analytical Center of the Russian Government (Open Data Bulletin, 2015-2016), the Annual report of the Russian NGO Information Culture (Begtin, Vovk, Sakoyan, 2015), government documents around open data (methodological recommendations for government bodies in data publishing (OGD Recommendations, 2014), the Plan of Open Data Russian Federation Development 2015-2016 (Russian OGD Plan, 2014), research reports of the Higher School of Economics (HSE, 2012), (Zhulin, 2013), information from the Federal OGD Portal of the R.F., research reports of the Infometr project, and reviews of Russian Federation Open Data Council working sections. Also, we used the questionnaire from (Ubaldi, 2013) to conduct surveys of government stuff developing the government OGD portals, extracting additional information. Three surveys have been conducted: one top manager, one data manager, and one technical leader (software architect).

## 5. RESULTS

A quick overview of the R.F. OGD ranking and OGD portals is in place before analysing the Russian OGD initiatives.

Table 1 shows how Open Global Data Index (OGDI, 2016) and Open Data Barometer (ODB, 2016) ranked the OGD in the R.F. In both R.F. place decreases from 2013 to 2015 that indicates higher OGD activity in R.F. on 2013. In OGDI the score of R.F. decreases from 2014 to 2015, but in ODB it slightly increases from 2013 to 2015. Difference results indicate various viewpoints under estimation of OGD, which used by OGDI and ODB. It should be noted that in 2014-2015 the R.F. spent most efforts to improve OGD quality and infrastructure rather than to increase quantitative metrics, and this is one of the reason decreasing Russian OGD indexes.

Table 1: OGD of R.F. in Open Global Data Index and Open Data Barometer.

| Year | Place | Score |
|---|---|---|
| Open Global Data Index | | |
| 2015 | 61 | 30% |
| 2014 | 45 | 43% |
| 2013 | 32 | 43% |
| Open Data Barometer | | |
| 2015 | 26 | 48,25% |
| 2013 | 20 | 44,79 % |

To consolidate OGD initiatives, the OGD portal of the R.F. (http://www.data.gov.ru) was launched in 2014. The structure of the portal follows the G8 Open Data Charter (G8, 2013) including a classification of datasets into 16 categories: Government, Economics, Education, Health, Culture, Ecology, Transport, Sport, Construction, Entertainment, Tourism, Electronic, Trade, Cartography, Security, and Weather. In total the portal contained about 7500 datasets in May 2016. On average, there were 8000 visits to the portal per month. The biggest contributors were the Federal Agency of Statistics, the Government of the Tula Region, the Moscow Government, the Government of Tomsk Region, and the Zelonograd Municipality. Statistics of data publishing in 2014-2016 is shown on fig. 1.

In addition to the federal portal there are a number other open data portals in the R.F. Examples of these are the Federation Spending Portal (http://zakupki.gov.ru/, launched in 2011) and the Federation Budget Portal (http://budget.gov.ru/, launched in 2013). These portals contain data in machine-readable formats with tools for data visualization and browsing. In 2014, Russian OGD on Government Spending ranked third in the world (OGDI, 2015). The R.F. is divided into 85 regions, and every region has an administration (regional government). Every regional government has its own web-resources, including OGD: 13 regional governments have ODG portals, and 29 ones have pages with OGD. The leaders are the Tula Region, Moscow, St Petersburg and the Ulyanovsk Region. Federal bodies and municipalities have the most OGD pages. Synchronization of federal and other OGDs is carried out automatically: all data from the regional portals and OGD pages of the various government bodies are copied/updated to the OGD Portal of the R.F. on the regular bases. Corresponding service was launched in the beginning of 2016, which led to a dramatic increase of the datasets on the federal portal as shown on fig. 1.

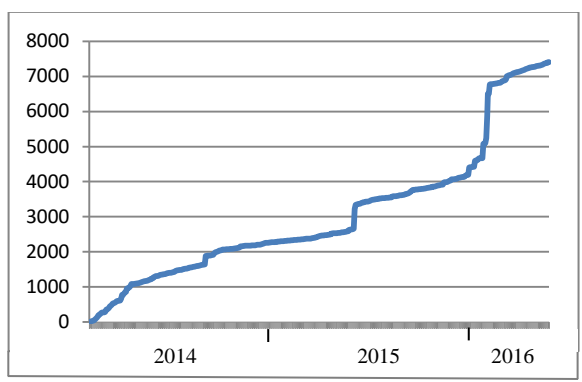

Figure 1: Changes in the number of datasets on the OGD portal of the R.F.

## 5.1 Overall vision of OGD in the Russian Federation

E-government and Open Government in the R.F. started with the administrative government reform in 2005. One of the main tasks of the reform was to develop a basis for free access to government information and to establish the legal and organizational framework for e-government services. Consequently, a number of government information systems to facilitate the information exchange for different government bodies and a network of e-government services portals were created in 2009 (Barabashev, Straussman, 2007), (Koznov, Chevzova, Samochadin, Azarskov, 2011). In late 2011, the Open Government project was launched in the R.F. The main purpose of the project was to provide communication between the federal government and various political parties, local and municipal authorities, civil society institutions, and common citizens. Among various activities it should be mentioned the launching portal of the R.F. (http://government.ru). The portal includes information about the federal government, all of the ministries, and regional governments. The monthly visitor count of the public services portal range between 200,000 and 700,000. As a result, according to the United Nation E-Government Survey 2012 (UN Report, 2014), the R.F. was one of the emerging leaders in e-government development in the world (7th place), advancing 32 positions from 2010 to 2012 in the world ranking of the United Nations (27th place in the general world ranking).

In the R.F. the OGD is a tool to implement the Open Government. This started in 2012 when the first official OGD concept was developed, and in 2013-2015 a number of laws and regulations were approved in the R.F. in order to support the OGD implementation. OGD development is also one of the main focus of ICT development in the R.F. according to the national ICT road map (ICT Road-Map, 2013). This document defines OGD as a tool to create e-services in the social sphere.

The coordination of OGD initiatives between central and local level is implemented as three level schema: central level, regional level, and region authority level. On central level laws, regulations, frameworks, programs, and recommendations are developed, and existing OGD is monitored and analysed. Building on this base regional governments create local legislation and programs. Various regional and municipal authorities follow these programs (regional authority level). Regional governments employ two general practices. Within the more efficient one, all regional government organizations load their data on the integrated regional government portal, and thus all OGD work is centrally-managed. The practice is implemented in St Petersburg, where more than 40 government organizations are subordinate to the St. Petersburg government. The other practice is that of independent work with data in each government organization in a region. The practice decreases the expenses of the regional government, but the quality of the data is lower.

## 5.2 Governance/institutional framework

In 2014, a program for the openness of the government bodies was prepared by the Russian government. It is aimed not only at making the information about government bodies public and accessible, but also at raising the efficiency of communication between the government and the citizens in order to improve the quality of administration, as well as to create tools to measure the openness of government bodies. The program is the general context of the OGD movement in the R.F.

In 2014, Russian Open Data Council developed the open data plan for the years 2015-2016 (Russian OGD Plan, 2014). The plan contains the list of expected actions with result assessment methods: the provision of methods of monitoring and using OGD, the improvement of legal support for OGD, the development of the OGD portal of the R.F., detecting new information to be represented as OGD, OGD ecosystem development, and involving NGOs and business companies into the OGD movement. In 2014, the Russian Government approved the road map of IT sector development for the years 2014-2025 (ICT Road-Map, 2013), and OGD development is one of the main priorities of the plan.

In 2014-2015, the «Methodological recommendations for publishing open data for government bodies» were developed to provide guidelines for government bodies in OGD activities (OGD Recommendations, 2014). The recommendations include requirements for licensing, determining the mandatory procedures for data publication, the rules for data publishing, data formats (CSV, XML, JSON, RDF), metadata format, and some other technical requirements.

The following organizations are working to coordinate and develop OGD at the federal level: the Russian Open Data Council, the Ministry of Economic Development, Ministry of Telecommunication, and the Government Analytical Center of the Russian Federation. The Open Data Council is a part of the Government Commission on Open Government; it coordinates the development of OGD through preparing government programs, proposals and recommendations, collecting and applying the best practices, promoting the idea of OGD, and through creating independent feedback channels. The Ministry of Economic Development is responsible for developing the federal portal, providing operational and procedural support and synchronization of federal and regional initiatives. The Ministry of Telecommunications is responsible for coordination of OGD-development by government bodies including corresponding information systems, as well as the advancement fo social e-services based on OGD. The Government Analytical Center of the Russian Federation monitors the OGD in the R.F.: the results of the analysis are published in Open Data Bulletin (Open Data Bulletin, 2015-2016) that has been issued every three months since June 2015.

## 5.3 Legal framework and policy environment

This section provides an overview of the most important legislation for open data in the R.F. The Federal Law 149 "On information, information technologies and information protection" (2006) and Federal Law 8 "On providing access to the information on the activities of governmental and municipal authorities" (2009) define the rights for information search, access, and transfer as well as the citizens' rights to access government information. The Russian Government Order No 953 adopted in 2009 determines and classifies the information that government bodies are to publish on the Internet, and prescribes update procedures for each information category. Presidential decree No 601 "On the general policy for improving government administration" issued on May 7, 2012 involves the figures for public enquiries handling and a roadmap for opening government information. The term "open data" was officially defined in Federal Law 112 (2013), which formed the legal basis for the government's work with open data. The Russian Government Resolution No 1187-r (July 2013) obliges Russian government authorities in federal, regional, and municipal levels to publish their data on the Internet and designates the types of information to be published in accordance with the Open Data Charter (G8, 2013). (Russian OGD Plan, 2014) states that by 2015 there will be a legal framework for open data, which, however, needs some revision and adaptation.

In 2014, some amendments to existing legislation were introduced (Federal law 35). They concern the use of open licenses in the R.F. which are equivalent to the licenses of the Creative Commons and GNU FDL. Licensing is a highly contentious issue for the country's OGD. In 2014, the OGDI indicated the lowest score in licensing for all OGD categories in the R.F. (OGDI, 2015). As of today, each dataset of the portal is supplied with a brief permission note granting the right to use it freely in any "appropriate, lawful purpose." The recommendations (OGD Recommendations, 2014) require that all data be published with a license based on free license. The text prescribes the content of the license, which conforms to (Open Definition, 2016). It is also said that the Creative Commons/Open Data Commons license could be used as a major guideline for licensing government data.

## 5.4 Technical issues

The technical issues of data opening in the R.F. follow (OGD Recommendations, 2014). Implementation oversight is carried out by the Open Data Council.

The quality of the data is monitored primarily in terms of the published data updates. Hence, according to the Russian Centre for Information and Analysis, only 30.7% of OGD was up to date as of mid-2015, while the leaders in data publishing on average had less than a half of their data updated. Based on (Open Data Bulletin, 2015-2016), only 26% of the datasets on the OGD Portal of the R.F. were up-to-date as of late 2015.

(OGD Recommendations, 2014) state, that the OGD published by the Russian government bodies are to have one of the following formats: CSV, XML, JSON, or RDF. Data on OGD Portal of the Russian Federation is presented on the following formats: CSV (63%); XML (36%); ZIP/GZ, JSON, XLSX/XLS and RDF (1%). The linking of data is very poor so far: the RDF-formal is used properly only by the Tula Region and the Ministry of Education portals.

Most of the Russian OGD-portals provide API (Application Programming Interface) for external programming data access. However, not all the data sets are accessible via APIs: according (Open Data Bulletin, 2015-2016) only 62% of data sets on federal OGD portal are available this way).

Most of Russian portals have built-in search engines that use key words, topics, data formats, organization names, and types. Some portals provide dataset visualization tools, but those are mostly limited to charts and tables.

As for the workflow for data release and approval, there is no common procedure, and each government body follows its own workflow. Open Data Council regular updates of mandatory publication list and carries out the management of publishing most demanded data.

Each portal has its own data storage. Due to relatively small amounts of data, there is no unified policy yet.

OGD portal of the R.F. is only one aggregator portal in the R.F. Obligatory metadata and their formats for all OGD officially published in the R.F. are prescribed in (OGD Recommendations, 2014).

## 5.5 Economic and financial

Open data development in the R.F. is financed by the government only. The Federal Portal is funded by the Ministry of Economic Development, while federal bodies, regional governments, and municipalities fund development of their OGD themselves. Business companies are not engaged in OGD funding.

## 5.6 Organizational issues

Let us now consider the measures undertaken in the R.F. to make changes in the public sector in relation with OGD.

Increasing the openness and accountability of government bodies is highly topical issue in the R.F. Thus, the Russian government openness program has been active since 2014. The program involves monitoring the openness of the government bodies by the non-government project Infometr based on a web-resource analyses. In December 2015, the Infometr project monitored the open data of the federal government bodies, checking them for compliance with the official requirements and plans. All the 77 federal government organizations were verified with a 55,1% average conformance with (OGD Recommendations, 2014).

Of the measures to shift the culture of the public sector towards OGD, the more noteworthy is the government program to promote open data awareness and popularity with officials (launched in 2015). The program intended to develop a number of education courses for civil servants. The Analytical Centre for the Government of the Russian Federation conducts educational webinars in OGD for the government staff. The Infometr project provides consulting in OGD to government organizations and staff. In 2014-2016, the Russian NGO Information Culture has run an education project titled "The School of Information Culture" to establish a dialogue between the government and the public through honing the skills in information exchange, sharing and processing. Some educational events on OGD are conducted by the NGO Committee for Civil Initiative.

It is of great importance for the public sector that their services for citizens can be improved with OGD through constructing new e-services. OGD portal of the R.F. presents 210 open data based software applications on different sectors: Tourism (47 applications), Government (46 applications), Transport (30 applications), Entertainment (24 applications), Culture (13 applications), etc. A considerable number of hackathons are held both federally and regionally to promote the development of OGD-based e-services.

## 5.7 Communication and interaction

The key actors of the Russian OGD ecosystem are:
- Government organization who is responsible for planning, implementing, and analysing OGD initiatives.
- Federal/regional government staff (analysts, managers and portal developers) who perform data collection and develop OGD websites.
- Software companies and developers who create e-services using OGD.
- NGOs and civil enthusiasts.
- Academic community.
- Citizens interested in OGD and OGD-based e-services.

To increase the public interest in OGD, a variety of measures are undertaken. The Higher School of Economics has carried out a number of research projects detecting business requests to OGD movement. Also, various events (Hackathons/Competitions, Conferences/Seminars) are conducted around OGD by government organizations and NGOs as shown in Table 2. It can be clearly seen that the NGOs are very active in that. Regional governments, such as St Petersburg, Ulyanovsk and some others, also conduct open data Hackathons/Competitions. A number of business companies are taking part on the OGD events: Yandex, Google (Russian Site), NextGIS and others. Russian academic community has organized a series of seminars, conferences and student schools on open data.

Non-government initiatives play an important role in Russia by facilitating and encouraging the public interest in OGD movement. The report (Begtin, Vovk, Sakoyan, 2015) describes other non-government OGD activities, below the most important organizations and initiatives are presented.
- NGO Information Culture (http://www.infoculture.ru/) conducts various events and projects to promote Open Government and OGD in Russia: a study of open data in Russia in 2015; the educational projects; the support of the growing Open Data Hub for all kinds of open data.
- Gis-Lab (http://gislab.info) is a community of experts in geo-information systems. In particular, the community normalizes and improves open geo-data from the portal of the Moscow government.
- Infometr (http://infometer.org/) is a non-government project for monitoring and assessing the quality of Russian government web-cites.
- The NGO Committee for Civil Initiative (https://komitetgi.ru) conducts various educational events on OGD.

Russian open data portals have started to collect user feedback on OGD published. In 2016, Information and Analytical Centre of Russian Government used user feedback as one of the metric to estimate quality of the OGD of various government organizations. The practice of opening datasets on user demand is also in use. The St. Petersburg government, for instance, has a positive experience of opening additional datasets on request. The OGD portal of the R.F. had 236 user requests in 2015, 70% were moderated, 24,6% were fulfilled (fully or in part). It must be said, though, that this is only the beginning.

Table 2: Open data events in Russia in 2011-2015.

| Type | Hackathons/ Competitions | Conferences/ Seminars |
|---|---|---|
| NGO | 10 | 11 |
| Gov (federal) | 6 | 4 |
| Gov (reg & municipal) | 11 | |
| Business | 2 | 4 |
| Universities | | 6 |
| Mass Media | 1 | |

## 5.8 Impact

Some members of the Open Data Council argue that the OGD initiative has already proved economically beneficial, although no precise figures have been presented so far. This part of the framework in the R.F. also calls for intensive development.

## 5.9 Challenges of OGD in the Russian Federation

The ODG movement as a cross-country initiative is facing challenges due to the large size of the Russian Federation (both territory and population). In addition, the OGD movement in the R.F. is quite recent: legislation concerning free access to government information was developed only in 2006-2015 (in Western countries similar acts and regulations started to be issued in the 1970s or even earlier); the federal OGD portal was launched in 2014 (both U.S. and U.K. – in 2010). It should also be noted that Russian civil society as such is truly young. Let us discuss the challenges in more detail following the components of the OCDE framework (Ubaldi, 2013).

*Overall vision.* One of the main OGD national priorities is the usage of OGD in new e-services development. This priority requires more systematic support in the current situation, when new e-services are being developed either by government organizations themselves, or by big software companies like Yandex (www.yandex.ru). On the other hand, many Hackathons have been conducted, yet they produce only prototypes rather than mature e-services. There is not enough support of small innovation software companies focusing on e-services based on OGD.

*Government/institutional framework.* OGD concepts, planning and guidelines need to be more detailed. In particular, different kinds of organizations should be identified from the OGD point of view. In particular, municipal and federal organizations have various "weight" and audiences, therefore, their data have different life cycles.

*Legal framework and policy environment.* Along with considerable progress in this area, licensing remains a problem as is highlighted in (Begtin, Vovk, Sakoyan, 2015). The existing permission notes (such as those on the OGD portal of the R.F.) do not qualify as licenses.

*Technical issues.* There are many problems with the quality and relevance of the published data. This is understood by the government and efforts are made to organize OGD monitoring (here we must note the Analytical Center of the Russian Government and the Infometr project). To overcome the existing problems, infrastructure measures are needed, which means, in particular, a closer collaboration with the bodies that perform the monitoring and the bodies that determine OGD polices.

*Business and economic.* It is necessary to provide detailed business, economic and financial models for OGD initiatives, and to stimulate business participation in the OGD movement. It will take the financial burden off the government in terms of developing OGD with the corresponding organizational and ICT infrastructure. The relations between the OGD movement and business in the R.F. are currently too weak to meet the needs of enterprises in real estate and insurance business, banking.

*Organizational issues.* Further efforts are required to shift the culture of the public sector towards OGD. Special measures are necessary to help public servants to find new opportunities in OGD. At the moment, they see OGD as a mandatory part of their work rather than a real working tool.

*Communication and interaction.* The most important issue is to implement the wide use of OGD. As (Hellberg, Hedström, 2015) shows, people generally seem to like the idea of open public data, but are not necessarily active in the data reuse process. This is equally true for the Russians, who need encouragement to use OGD.

*Impact.* Monitoring the activities around Russian OGD should be expanded from the data themselves to estimating the perspectives and assessing the results of OGD for the economy and society of the country.

# 6. CONCLUSIONS

In this paper, we tried to close the gap in systematic research of the OGD in the R.F. by conducting a study based on the OCDE framework (Ubaldi, 2013). The OGD movement in the R.F. has made considerable progress: a number of OGD portals have been implemented, the federal OGD portal as a data aggregator has been developed, and underlying government ICT and organisational infrastructure has been created and is constantly improving. The volume of OGD is increasingly growing, while measures to improve its quality and relevance are being undertaken. A number of new e-services based on OGD are being developed. The challenges Russia faces today can be overcome in the future if the OGD movement is implemented more systematically and becomes more integrated into the society. Moreover, the efficiency of Russia's OGD movement is tightly connected with the country's general progress, including economy, open government initiatives, and civil society.